\documentclass[referee]{aa} % for a referee version
\usepackage{graphicx}
\usepackage{txfonts}

\usepackage{natbib}
\usepackage{color}
\long\def\comment#1{}
\def\Dvect#1{\mbox{\boldmath $#1$}}

\def\NNnea{18,000}

\definecolor{purple0}{rgb}{0.625, 0.1250, 0.9375}
\definecolor{purple1}{rgb}{0.605, 0.1875, 1.0000}
\definecolor{purple2}{rgb}{0.569, 0.1725, 0.9333}
\definecolor{purple3}{rgb}{0.490, 0.1490, 0.8039}
\definecolor{purple4}{rgb}{0.333, 0.1020, 0.5451}
\definecolor{lightgreen}{rgb}{0.5647, 0.9333, 0.5647}
\definecolor{forestgreen}{cmyk}{0.91, 0.0, 0.88, 0.12}
\definecolor{pinegreen}{cmyk}{  0.92, 0.0, 0.59, 0.25}
\definecolor{olivegreen}{cmyk}{ 0.64, 0.0, 0.95, 0.40}

\def\mynotecolor{black}

\def\myfigwidthN{0.50}
\def\myfigwidthW{0.98}

\begin{document}

   \title{Asymmetric impacts of near-Earth asteroids on the Moon}
  %\subtitle{I. Overviewing the $\kappa$-mechanism}

   \author{Takashi Ito\inst{1}
          \and
          Renu Malhotra\inst{2}
%\fnmsep\thanks{Just to show the usage of the elements in the author field}
          }

   \institute{National Astronomical Observatory,
               Osawa 2--21--1, Mitaka, Tokyo 181--8588, Japan\\
              \email{ito.t@nao.ac.jp}
         \and
               Lunar \& Planetary Laboratory,
               University of Arizona,
               1629 E. University Blvd., Tucson, AZ 85721--0092, USA\\
%            \thanks{The university of heaven temporarily does not
%                    accept e-mails}
             }

   \date{Received ---, ---; accepted ---, ---}

% \abstract{}{}{}{}{} 
% 5 {} token are mandatory
 
  \abstract
  % context heading (optional)
  % {} leave it empty if necessary  
   {
Recent lunar crater studies have revealed an asymmetric
distribution of rayed craters on the lunar surface. 
   The asymmetry is related to the synchronous rotation of the Moon:
there is a higher density of rayed craters on the leading hemisphere
compared with the trailing hemisphere.  Rayed craters represent
generally the youngest impacts.}
  % aims heading (mandatory)
   {The purpose of this paper is to test the hypotheses that (i) the
population of Near-Earth asteroids (NEAs) is the source of the
impactors that have made the rayed craters, and (ii) that impacts by
this projectile population account quantitatively for the observed
asymmetry.}
  % methods heading (mandatory)
   {We carried out numerical simulations of the orbital evolution of a
large number of test particles representing NEAs in order to determine
directly their impact flux on the Moon. The simulations were done in
two stages. In the first stage we obtained encounter statistics of
NEAs on the Earth's activity sphere. In the second stage we calculated
the direct impact flux of the encountering particles on the surface of
the Moon; the latter calculations were confined within the activity
sphere of the Earth. A steady-state synthetic population of NEAs was
generated from a debiased orbital distribution of the known NEAs.}
  % results heading (mandatory)
   {We find that the near-Earth asteroids do have an asymmetry in
their impact flux on the Moon: apex-to-antapex ratio of
$1.32 \pm 0.01$.
   However, the observed rayed crater distribution's asymmetry is
significantly more pronounced: apex-to-antapex ratio of
$1.65 \pm 0.16$.
  Our results suggest the existence of an undetected population of slower
(low impact velocity) projectiles, such as a population of objects nearly
coorbiting with Earth; more observational study of young lunar craters is
needed to secure this conclusion.
}
  % conclusions heading (optional), leave it empty if necessary 
   {}


   \authorrunning{Ito and Malhotra}
   \titlerunning{Asymmetric cratering on the Moon}

   \maketitle
%
%________________________________________________________________

%%%%%%%%%%%%%%%%%%%%%%%%%%%%%%%%%%%%%%%%%%%%%%%%%%%%%%%%%%%%%%%%%%%%%%%%%%%%%%
\section{Introduction}

It is well known that many satellites of the solar system planets are
locked in synchronous rotation --- their mean rotational angular speed
and mean orbital motion is in a 1:1 commensurability.
The synchronous rotation of these satellites leads to asymmetric spatial
distribution of impact craters on these satellites: the leading hemisphere
tends to have more craters than the trailing hemisphere.  
Such leading/trailing asymmetries in crater distributions have been
observed on the Galilean satellites of Jupiter and on Neptune's moon
Triton \citep[e.g.][]{shoemaker82,schenk99,zahnle98}.

Such an asymmetry was recently confirmed on the Moon.
A detailed analysis of the Clementine 750-nm mosaic images has revealed that
there is spatial variation in the density of rayed craters on the Moon
\citep{morota2003}.
Lunar rayed craters are morphologically young and fresh craters with bright
rays, generally estimated to be younger than 0.8 billion years
old \cite[e.g.][]{mcewen97}.
A total of 222 rayed craters larger than 5 km in diameter $(D)$ are
identified in the study area of about $1.4 \times 10^7$ km${}^2$.
The average density of rayed craters on the leading side of the Moon
is found to be substantially higher than that on the trailing side, and
the observed ratio of crater density ($D > 5$ km) at the apex to
that at the antapex is about 1.65.
Based on a rough analytical estimate, \citet{morota2003}
conclude that this ratio suggests that recent craters
on the Moon are formed mainly by the near-Earth asteroids which have
lower impact velocities, rather than comets that have systematically
higher impact velocities. 
Whether or not these conclusions are correct, it is true
that the ratio of crater densities of the
leading side and the trailing side of the Moon contains a significant
amount of information about the type of projectile populations that
have created 
craters on the lunar surface over the past $\sim 1$ billion years and
under what kind of dynamical conditions.%

The origin of the leading/trailing asymmetry of impact craters on a
synchronously rotating 
planetary satellite is qualitatively explained as follows.  Assume
that the source of impacts 
is a heliocentric population of small objects on modestly eccentric
and inclined orbits. In the 
rest frame of the planet, this population appears almost isotropic and
the impact velocity 
vectors have an isotropic distribution.
The impact craters asymmetry occurs because the satellite in
synchronous rotation encounters 
projectiles more often on its leading side than on its trailing side.
Furthermore, the average impact velocity of projectiles tends to be
larger on the leading side than on the trailing side due to the difference
of average relative velocities between the projectiles that encounter
the leading and the trailing sides; this leads to systematically larger
craters on the leading hemisphere compared with the trailing hemisphere.

The degree of the asymmetric crater distribution is a function of
satellite's orbital velocity and the average relative velocity between
projectiles and the planet--satellite system.
When a satellite with synchronous rotation has a large orbital velocity
around its mother planet, or when the average relative velocity between
projectiles and the planet--satellite system is small,
the asymmetric distribution of craters becomes more pronounced.
Smaller orbital velocity of the satellite, or larger average
relative velocity of projectiles 
tends to diminish the asymmetry of crater distribution.

The purpose of the present paper is to quantitatively test the hypothesis that
impacts from the NEA population (with its currently known properties)
account for the observed asymmetric crater distribution on the Moon.
We do this by carrying out extensive numerical integrations of test
particles to simulate the impact flux of NEAs. 
In order to obtain impact statistics and impact velocity distribution,
%%% as accurately as possible,
we calculate direct impacts of projectiles
on the Moon without analytical approximation.
In Section \ref{sec:model} we describe our dynamical model, our numerical 
method and our choice of initial conditions, and a description of the
first stage of our numerical simulations in which we trace the
dynamical evolution of test particles from 
their initial locations to the edge of Earth's activity sphere; our results on 
 NEA encounters with
the Earth's activity sphere are given in section \ref{sec:EAS}.  
Section \ref{sec:statMoon} describes the second stage of our numerical
simulation in which we trace the evolution of particles within the
Earth's activity sphere to obtain impact fluxes, impact velocities and
their spatial distribution on the Moon. Section \ref{sec:comparison}
provides a comparison of our simulation results with the observations
of the lunar crater record. Section \ref{sec:discussion} is devoted to
discussion of the results, and Section \ref{sec:conclusions} to a short summary and conclusions.

While our work was in progress, \citet{gallant2009} published a study 
with quite a similar motivation to ours, which also yielded a similar
numerical result about the expected lunar cratering asymmetry from NEA impacts.
Although a large part of our results overlap, the numerical models are different.
As we describe in section 2, our numerical model is
%%% more
straightforward and
includes the orbit evolution of NEA-like particles from their current
orbits up to their impacts, while \citet{gallant2009}'s study uses the NEA orbits
without dynamical evolution. In this regard our paper serves as a 
complement to \citet{gallant2009}.  We also consider some additional important
implications of the results that were not discussed previously.

%%%%%%%%%%%%%%%%%%%%%%%%%%%%%%%%%%%%%%%%%%%%%%%%%%%%%%%%%%%%%%%%%%%%%%%%%%%%%%
\section{Numerical model\label{sec:model}}
\begin{figure*}
  \centering
  \includegraphics[width=\myfigwidthN\textwidth]{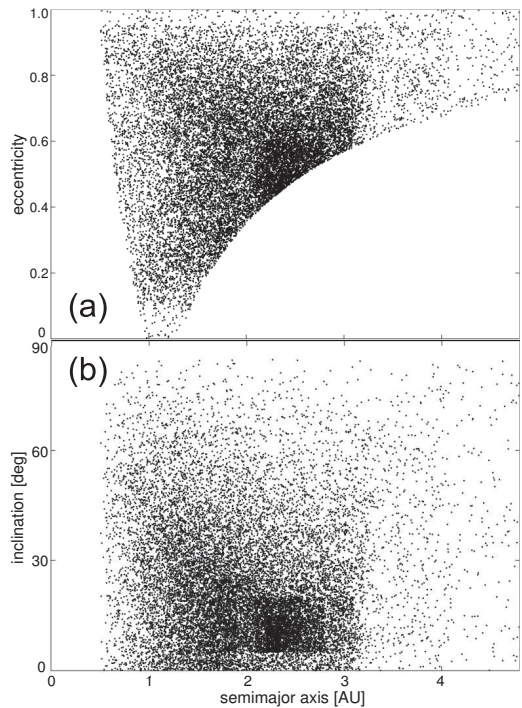}
  \caption{Initial osculating orbital elements of the 
           NEA-like particles in our numerical model.
  (a) Semimajor axis vs. eccentricity.
  (b) Semimajor axis vs. orbital inclination.
(Inclinations are referred to the ecliptic at J2000.0.)
}
  \label{fig:elem_ini}
\end{figure*}

Our numerical model comprises of two stages.
In the first stage, our numerical integrations include the eight major planets
and the Sun, and a large steady-state number of test particles with NEA-like orbits  (Fig.~\ref{fig:elem_ini}). 
We numerically integrate their orbital evolution for up to 100 million years.
Throughout these integrations, we record all close encounters of the particles 
that reach the Earth's activity sphere. 
(Note that in the first stage of calculation the Moon is not included.)
We use this record in our second stage of numerical simulation, in which we adopt the restricted
$N$-body
model consisting of the Earth, the Moon, and the Sun, and cloned test particles
within the Earth's activity sphere (as described in detail in Section
\ref{sec:statMoon}). In the second stage, we do not include the effects
of any planets save the Earth but we include the Moon's gravity.

%%%%%
%%%%% Figure 1. Initial (a,e,i) distributions
%%%%%

For our first stage numerical simulation we generated a synthetic population of particles
with orbital 
elements similar to the ``debiased" distributions of near-Earth objects 
(NEOs) described in \citet{bottke2002}.
(Note that NEOs are largely composed of NEAs,
so we will keep using the term NEAs rather than NEOs in this paper.)
Specifically, we generated {\NNnea} particle initial conditions whose
distributions of $a$, $e$, and $I$ obey the histograms
shown in Figure 12 of \citet{bottke2002} which gives the debiased orbital 
distribution of the near-Earth asteroids of absolute magnitude $H<18$.
%%%
\footnote{%
Note that at this stage of our calculation we do not consider at all the
size-frequency distribution (or absolute magnitude distribution) of the
particles.%
}
The orbital elements of our synthetic NEA population are shown in
Fig.~\ref{fig:elem_ini}.
This population represents a good
snapshot of current orbital distribution of NEAs.
Studies of impact craters in the inner solar system
indicate that there has been a relatively constant supply of
impactors over the past three billion years
which has kept the impactor flux around the Earth--Moon system close to
a stationary state  \citep{mcewen97,ivanov2002}, and that this impactor
population is dominated by NEAs \citep{strom2005}.

For the numerical integration scheme we used
the regularized mixed-variable symplectic method \citep{levison94}.
The basic framework of our first stage simulation follows \citet{ito2006}.  
When a test particle approaches within the physical radius of the Sun
or that of planets, we consider the particle to have collided with
that body and lost from the NEA population.
Also, when the heliocentric distance of a test particle exceeds
 100 AU, the particle is considered lost.  
Over the 100 Myr length of the simulation, a large fraction ($\sim90\%$, e.g.~\citet{ito2006}) of
the synthetic population would be expected to be removed in this way,
and if this loss were not compensated, we would not be able to
mimic a steady-state NEA flux. 
We realize the steady-state NEA flux in our numerical simulation
as follows: for each ``lost'' particle, we immediately introduce in our
simulation another particle with the original position and velocity of
that ``lost'' particle.
% \textcolor{\mytextcolor}{%
% For example, when particle $i$ is removed from the simulation
% by any of the reasons described above
% at the position $\Dvect{r}_i$ and the velocity $\Dvect{v}_i$,
% another particle, also denoted by the subscript $i$, with the position
% $\Dvect{r}_{i,0}$ and the velocity $\Dvect{v}_{i,0}$ is immediately
% introduced in the simulation where $\Dvect{r}_{i,0}$ and $\Dvect{v}_{i,0}$
% are the initial position and the velocity of the particle $i$ at the
% beginning of the integration.  
% }
This procedure achieves a steady-state population
of NEAs in our simulation.  In particular, we verified that the
distribution of lunar impact velocity remains steady throughout the
simulation timespan.

%%%%%%%%%%%%%%%%%%%%%%%%%%%%%%%%%%%%%%%%%%%%%%%%%%%%%%%%%%%%%%%%%%%%%%%%%%%%
\subsection{Particle encounters with Earth's activity sphere\label{sec:EAS}}

%%%%%
%%%%% Figure 2. Statistics at rI
%%%%%
\begin{figure*}
  \centering
  \includegraphics[width=\myfigwidthN\textwidth]{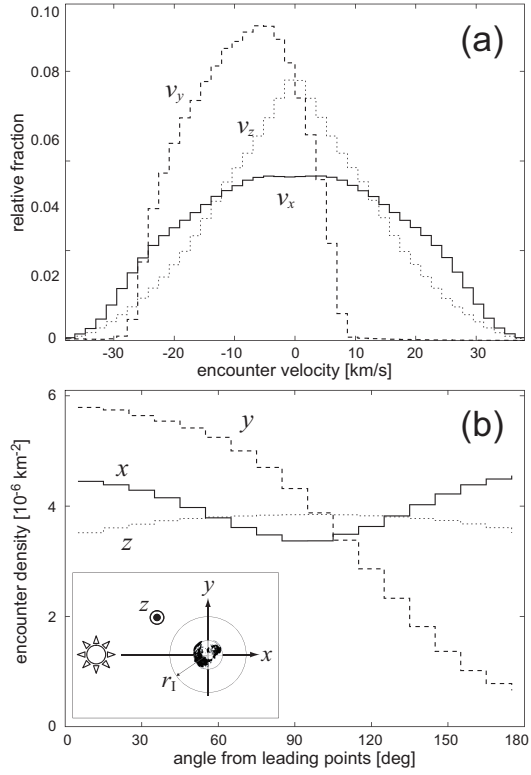}
  \caption{
         Statistics of NEA-like particle encounters
             at the Earth's activity sphere, $r_{\rm I}$.
             (a) Distribution of encounter velocity components $(v_x, v_y, v_z)$;
         The horizontal unit is km/s. 
     (b) Surface density of encounters at the surface of the Earth's
         activity sphere as a function of the angle from the leading points
         of each of the $x$, $y$, and $z$ directions.
         The vertical unit is $10^{-6}$ ${\rm km}^{-2}$.
 The inset in the panel b schematically shows
         the coordinate system $(x,y,z)$ adopted for this figure:
         The Sun always lies along the $-x$ direction,
         $+z$ is the normal to the Earth's orbit,
         the Earth practically goes toward $+y$ direction, and
         $r_{\rm I}$ is the radius of Earth's activity sphere.
         Coordinate of the leading point
         for the $y$ data in the panel is $(0, r_{\rm I}, 0)$.
}
  \label{fig:rI_stat}
\end{figure*}

Over the 100 myr simulation of a steady-state swarm of 18,000 particles,
we found 3,998 collisions with Earth.
%\textcolor{\mynotecolor}{%
%(Note that in our numerical model same particles hit the Earth many times
%due to how we mimic the steady-state flux of the NEA-like particles.)
%}
% This is clearly not a large number;
% based on a ratio of $\sim$1/20 for the geometrical collision cross-section of
% Earth-to-Moon,
% it is clear that only $\sim$200 collisions would result on the Moon,
% a number that 
% would provide impact statistics too poor to discern a leading/trailing
% asymmetry for useful comparison with the observed lunar crater record.
% This is the motivation for our second stage of simulation which we describe
% in the next section.
% 
We note that, although the number of planetary collisions is not large in
our first stage numerical integrations, there are many more encounters
at the planetary activity sphere of Earth.
The activity sphere, also known as the sphere of influence,
has a radius of $(m/M)^{2/5} d$ where
$m$ is the mass of a planet, 
$M$ is the mass of the Sun, and
$d$ is the heliocentric distance of the planet \citep{danby92}.
Earth's activity sphere, hereafter denoted $r_{\rm I}$, 
is about 144 times as large as the Earth's radius.
In our first stage numerical simulation, we recorded the encounters of
particles at the Earth's activity sphere over the 100 Myr integration,
and found
% 6,507,298 encounters of the population A,
% 5,395,606 encounters of the population B,
42,099,969 encounters.
This number is large enough to establish a time-dependent orbital
distribution function of the particles,  $F(a,e,I,\omega,\Omega,l; t)$
that can be used to create ``clones'' of particles in order to
increase the reliability of the collision statistics between the
particles and the Earth or the Moon, as we describe in the next section. 

In Fig.~\ref{fig:rI_stat} we show time-integrated distribution 
(over the 100 Myr duration of our simulation) 
of encounter velocity components and encounter density (number of
encounters per unit surface area) at the Earth's activity sphere
of the particles. 
In our simulation the average encounter velocity of the particles
at the Earth's activity sphere is $22.5$ km/s.
%%%
%%% more accurately, 22.48
%%%
We show these distributions
along all three spatial directions, $x$, $y$ and $z$. There are several
noteworthy features in these distributions.  While the encounter
velocity distributions with respect to the $x$ and $z$ directions are
symmetric, the $y$-direction distributions are markedly asymmetric.
Because of its very small orbital eccentricity,
the Earth's orbital motion is practically along the $+y$ direction.
We see the expected effect that more particles encounter the Earth's
activity sphere on its leading side (from the positive $y$ direction)
than its trailing side: the fraction of encountered particles having
negative $v_y$ is larger than that of the particles having positive
$v_y$ (the panel (a)). Consistent with this, we see in the panel (b)
that the encounter density is higher on the leading
hemisphere ($0-90^\circ$ angle with respect to the leading point
in the $y$ direction), and lower on the trailing hemisphere.

In the panel (b), we also notice a pattern in
the encounter density distributions along the $x$ direction:
the encounter density of the particles is somewhat
higher around the angle $\sim 90^\circ$ (with respect to the leading point
of the $x$ direction, $(r_{\rm I}, 0, 0)$).
The pattern of distribution in the $x$-direction is understood by the
following consideration. Because of its very small orbital eccentricity,
the Earth's orbital motion is practically along the $+y$ direction.
Along the $y$ axis, the angle with respect to the $x$- leading point,
$(r_{\rm I}, 0, 0)$, is $90^\circ$; this defines the solar terminator at the 
Earth's activity sphere.
Particles that encounter Earth in this region have very small
velocity relative to Earth, particularly when their random orbital 
velocity is low; the lower 
average encounter velocity leads to the lower encounter frequency of 
particles in this region.  This accounts for the dip near angle $90^\circ$
for the $x$ curves in the panel (a). Moreover, this effect would get smaller
when the particle population has larger random orbital velocity.

We also notice a pattern in the encounter density distributions along the $z$
direction in Fig.~\ref{fig:rI_stat}(b).
However, the dips near $90^\circ$ are of noticeably smaller magnitude
in this direction than in the $x$ direction, and the sign is the opposite:
we see a concentration of encounters around $90^\circ$.
We note that the vertical scale height of the NEA-like particle population,
approximately given by the average $\left< a \tan I \right>$,
is much larger than the radius of the Earth's activity sphere
($r_{\rm I} \sim 0.006$ AU).
This is why the encounter frequency of the particles does not 
vary as much along the $z$ direction compared with the $x$ direction.

From the above results on the distribution of particle encounters on
the Earth's 
activity sphere shown in Fig.~\ref{fig:rI_stat}, it is clear that the Earth
receives more NEA impacts on the leading (positive $y$) hemisphere than on
the trailing hemisphere.
This asymmetry leads to AM/PM asymmetries in NEA impact events,
which is discussed in \citet{gallant2009} in detail.

%%%%%%%%%%%%%%%%%%%%%%%%%%%%%%%%%%%%%%%%%%%%%%%%%%%%%%%%%%%%%%%%%%%%%%%%%%%%%%
\section{Asymmetric impacts on the Moon\label{sec:statMoon}}

Using the particle encounters at Earth's activity sphere,
we generated cloned particles by perturbing the encounter position 
$\Dvect{r}$ and velocity $\Dvect{v}$ of each of the original particles
so that their initial trajectories at the activity sphere become
slightly different:
$\Dvect{r}_{\rm clone} = (1+\delta_r) \Dvect{r}_{\rm original}$ and
$\Dvect{v}_{\rm clone} = (1+\delta_v) \Dvect{v}_{\rm original}$, where
$\delta_r$ and $ \delta_v$ are random numbers in the range
$[-0.1,0.1]$.
This procedure produces
a large number of particles that obey nearly the same orbital distribution
function as the original particles (``$F$'' in the previous
description, see Fig.~\ref{fig:rI_stat}) but with somewhat different
paths toward the Earth (and the Moon).

We repeated this cloning procedure five hundred times from the result
of the first stage numerical integrations,
generating 21.049895 billion particle initial conditions
on the Earth's activity sphere.
Using these sets of cloned particles, we performed a
second set of numerical integrations, this time with the restricted
$N$-body
problem including the Sun, the Earth, the Moon, and the cloned test particles.
Here we did not include the effect of other planets than the Earth, but we
included the Moon's gravity. All the cloned particles started near the Earth's
activity sphere, and were integrated until they hit
the Earth or the Moon or went out of the sphere.
We used the present orbital elements of the Moon with true anomaly
randomly chosen from 0 to $360^\circ$ for each of the 500 sets of clones.
We employed the regularized mixed-variable 
symplectic method again with a stepsize of 84.375 seconds (= $2^{-10}$ days).
(This small step size was arrived at by a process of trial to ensure that
even high velocity particle collisions with the Moon were not missed.)
Calculations were carried out in the geocentric frame.

%%%%%
%%%%% Figure 3. Impact statistics on the Moon
%%%%%
\begin{figure*}
  \centering
  \includegraphics[width=\myfigwidthN\textwidth]{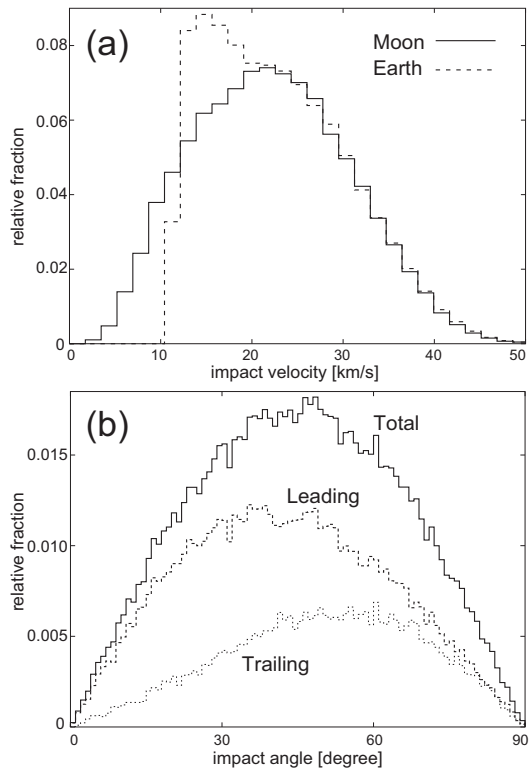}
  \caption{Statistics of impactors on the Moon and on the Earth.
  (a) Distribution of impact velocity on the Moon (solid line)
      and on the Earth (dashed line) of the clones.
  (b) Distribution of impact angle on the Moon on the entire surface
      (solid line; denoted as ``Total''),
      on the leading hemisphere (dashed line; denoted as ``Leading''), and
      on the trailing hemisphere (dotted line; denoted as ``Trailing'').
  }
  \label{fig:rM_stat1}
\end{figure*}

The second stage numerical integrations yielded
1,509,364 collisions with the Earth and
   73,923 collisions with the Moon.
Fig.~\ref{fig:rM_stat1} shows the distribution of impact velocities and
impact angles on the Earth and on the Moon.
Overall, the average impact velocities of the clones on the lunar surface,
($\sim 22.4$ km/s) is almost the same as
the average encounter velocity of the original particles
at the Earth's activity sphere.
%
%  dset   v_rI_avr v_rI_rms  | vM_imp_avr vM_imp_rms
% ---------------------------+----------------------
%  y(A)   18.52    20.30     | 18.46      20.25
%  Y(B)   23.32    24.83     | 23.24      24.75
%  U(B')  22.42    24.13     | 22.41      24.16
%
This means that lunar gravity plays only a minor role in accelerating
particles to the lunar surface in our numerical model.
Not only lunar gravity but the Earth's gravity also plays only a small role:
average impact velocity of the clones at the Earth's surface is 
$\sim 23.1$ km/s,
%%% accurately, 23.09 km/s
not being very different from the average impact velocity
with the lunar surface, in spite of the large difference
of the escape velocities from the two bodies
($\sim 11.2$ km/s on the Earth and $\sim 2.4$ km/s on the Moon).
The ratio of the number of collisions
with the Earth and those with the Moon is found to be
$20.4 \pm 0.1$.
%%% accurately, 20.42 = (1509364/73923)
For comparison, we note that \citet{zahnle97} reported the ratio of 
collisional cross sections of the Earth and the Moon as $\sim 23$, 
by assuming isotropic collisions and
average impact velocity of Earth-crossing asteroids 
to be 16.1 km/s on the Earth.

%
% On the impact angle distribution
%
Regarding the impact angles on the lunar surface,
we note that from simple geometrical considerations for an isotropic 
distribution of impact direction, the impact angle distribution is
expected to have a probability density function proportional
to $\sin 2i$ \citep{shoemaker62}
where $i$ is impact angle and $i \to 0$ means oblique impact.
The results of our simulation, taken all together, are consistent
with isotropic impact angles (Fig.~\ref{fig:rM_stat1}b).
We can mention in passing that there is a small but noticeable
difference in the impact angle distribution 
on the leading and trailing hemispheres of the Moon: the trailing hemisphere
slightly disfavors oblique impacts whereas oblique impacts are
slightly enhanced on the leading hemisphere, Fig.~\ref{fig:rM_stat1}(b).
This is not of significance
for the statistical results in the present paper, but it may be of interest
for future studies of individual lunar craters.

\begin{figure*}
  \centering
  \includegraphics[width=\myfigwidthN\textwidth]{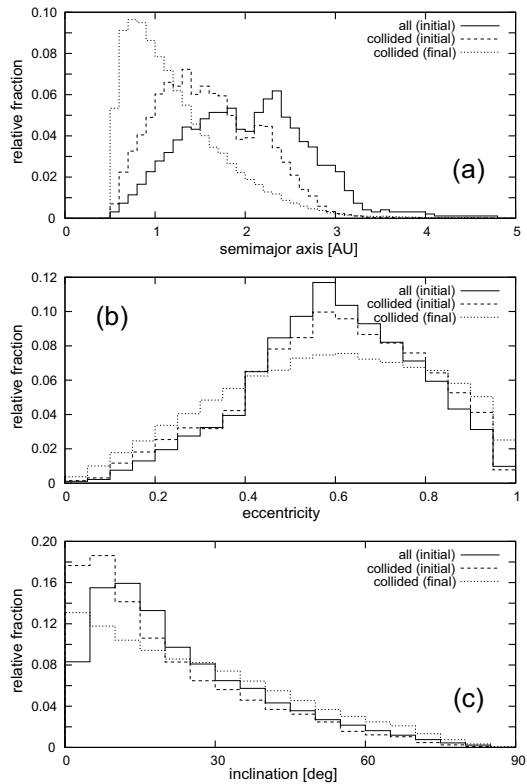}
  \caption{
      Heliocentric orbital distribution of
      (a) Initial semimajor axis distribution of
          all particles included in the first stage integrations (solid line),
          initial semimajor axis distribution of
          the particles (their equivalent clones) that
            eventually collided with the Moon (dotted line),
          and
          final semimajor axis distribution of
          the particles (their equivalent clones) that
            eventually collided with the Moon
           (dashed line; i.e. when they hit the lunar surface).
          Note that the dip around 2 AU in the solid line distribution
          is originated from the debiased semimajor axis distribution
          of \protect{\citet{bottke2002}}.
      (b) Same as (a), but for eccentricity.
      (c) Same as (a) and (b), but for orbital inclination.
  }
  \label{fig:dist_elem}
\end{figure*}

It is also interesting to examine the orbital element distribution
of the lunar colliders.  In Fig.~\ref{fig:dist_elem}, we plot histograms 
of the distribution of $a,e,I$ for the lunar colliders as well as for our synthetic
NEA initial conditions; for the lunar colliders, we plot histograms of both
their initial orbits and their orbits just before impact with the Moon.
Comparison of the two initial orbit distributions shows that
the lunar collider population has a higher fraction of low inclinations and 
low semimajor axes compared to the overall NEA initial orbit distribution.
Comparison of the initial orbits and final (just before impact) orbits
of the lunar colliders shows that there is significant dynamical evolution of
orbital elements that occurs prior to lunar impact: on average, semimajor axes 
evolve to lower values, eccentricities and inclinations evolve to higher
values.
% This dynamical evolution is undoubtedly owed to the many close encounters 
% with Earth-Moon that occur prior to the final collision.
This evolution takes place during the first several million years of
their trajectories.

%%%%%
%%%%% Figure 4. Orbital distribution of the collided particles with the Moon
%%%%%

%%%%%%%%%%%%%%%%%%%%%%%%%%%%%%%%%%%%%%%%%%%%%%%%%%%%%%%%%%%%%%%%%%%%%%%%%%%%
\section{Simulation compared with lunar crater data\label{sec:comparison}}

The second stage of our numerical simulation yields the spatial 
distribution of NEA impacts on the lunar surface.  
As mentioned in the introductory section,
in order to compare the distribution of impacts in our numerical model
with the actual lunar crater record, we have to consider a correction to the
raw numerical results due to the systematic difference in the impact velocities
on the leading and trailing hemispheres, a difference
that owes to the orbital motion of the satellite about its mother planet.  
This correction, which turns out to be quite small for the Moon,
arises as follows.

\begin{figure*}
  \centering
  \includegraphics[width=\myfigwidthN\textwidth]{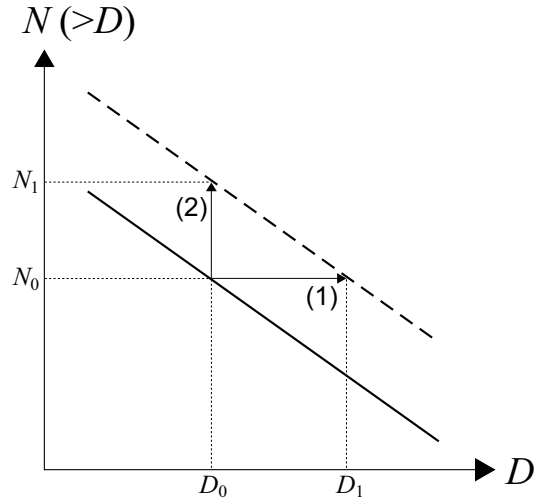}
  \caption{Schematic illustration of the change of
       crater SFD due to apparent change of crater size.
       $D$ is crater diameter, and $N(>D)$ is the cumulative
       number of craters whose diameter is greater than $D$.
       Adopted and modified from \protect{\citet{ishizaki97e}}.
  }
  \label{fig:N_D}
\end{figure*}

For a satellite with synchronous rotation,
the average impact velocity of projectiles is
somewhat larger on the leading side than on the trailing side.
This difference means that, on average, the apparent crater size would be
larger on the leading side than on the trailing side (assuming the projectile
size-frequency distribution (SFD) is not different on the two sides).
To illustrate the effect this has on the crater densities,
consider a power law SFD of craters, as in the solid line shown 
schematically in
Fig.~\ref{fig:N_D} where $N$ is the cumulative number of craters per unit area.
As a result of the higher (lower) average impact velocity on
the leading (trailing) side, 
the SFD curve of the craters on the leading (trailing) side becomes shifted
toward the positive (negative) direction along the horizontal $(D)$
axis, as indicated by the arrow {\sf (1)} in the figure.
This horizontal shift is practically equivalent to a vertical shift of the
SFD curve toward the dotted line in Fig.~\ref{fig:N_D}, as indicated
by the arrow {\sf (2)} in the figure, 
illustrating that $N$ gets larger on the leading side (smaller on the
trailing side) for the entire range of crater diameter, $D$.

The magnitude of the shift depends upon the relationship between the
impact velocity $v_{\rm imp}$ and the crater size, $D$.
Here we employ the Pi-group scaling
\citep{schmidt87,melosh89,housen91} where approximately
  $D \propto v_{\rm imp}^{\alpha}$ with $\alpha = 0.44$.
For the cumulative SFD of craters, we adopt%
\footnote{%
This SFD is also consistent with the young crater populations linked to
impacts by the NEA population on all terrestrial planets and the Moon
\citep{strom2005}.%
}
$N (>D) \propto D^{\beta}$ with $\beta = -2$
which represents well the SFD of the young rayed craters \citep{morota2003}.
When the average impact velocity is changed from $v_0$ to $v_1$, the
cumulative number density of craters at any given diameter $D$ changes
from $N_0$ to $N_1=N_0(v_1/v_0)^{- \alpha \beta}$.
This holds for any values of $D$ as long as the crater SFD is expressed
by a single power law.

%%%%%
%%%%% Figure 5. Schematic N-D relationship
%%%%%

From the results of our second stage simulation, we computed
the average impact velocity, $\langle v_{\rm imp}\rangle$ in km/s,
of NEAs on the lunar surface
as a function of angle from apex, $\gamma$ (degrees), by a least squares fit. 
We find
% $\langle v_{\rm imp}\rangle = -0.00245 \gamma + 18.65$ for population A,
% $\langle v_{\rm imp}\rangle = -0.00543 \gamma + 23.67$ for population B, and
$\langle v_{\rm imp}\rangle = -0.00672 \gamma + 22.7$.
                                %%% accurately, 22.67
This indicates 
that difference of $\left< v_{\rm imp} \right>$ between
the $\gamma = 90^\circ$ point and the
apex ($\gamma = 0$) or antapex $(\gamma = 180^\circ)$ is
% less than 0.22 km/s for the population A,       % 0.00245*90
% less than 0.49 km/s for the population B, and   % 0.00776*90
less than 0.61 km/s.                              % 0.00672*90
Compared with the average of $v_{\rm imp}$ over the entire range of
$0 \leq \gamma \leq 180^\circ$,
these velocity differences amount to
% $\lesssim 1.19{\%}$ for the population A,
% $\lesssim 2.10{\%}$ for the population B, and
$\lesssim 2.71{\%}$,
thus corresponding corrections to the cumulative SFD are given by
% $N_1/N_0 \sim 1.0105$,
% $N_1/N_0 \sim 1.0185$, and
$N_1/N_0 \sim 1.02$.
%%% accurately, 1.0238
This small difference is owed to the fact that the lunar orbital velocity of
$\sim 1$ km/s is much lower than the average impact velocity
$\langle v_{\rm imp}\rangle$ of $\sim 22$ km/s.
As a result of this small dependence, apparent change of the crater
SFD due to the impact velocity difference between the leading side and the
trailing side is quite modest.
This effect, however, would be important when considering the
asymmetric crater distribution on a satellite with higher orbital
velocity around its mother planet such as Ganymede around Jupiter 
\citep{shoemaker82,zahnle2001}.

%%%%%%%%%%%%%%%%%%%%%%%%%%%%%%%%%%%%%%%%%%%%%%%%
%\begin{quote}
%  \textcolor{red}{Renu, about the original Fig.~6: 
%  I think we can drop this figure. The text I added in the above paragraph
%  suffices for the purpose here.}
%  \par
%  \textcolor{\mytextcolor}{Takashi: Agreed.}
%\end{quote}
%%%%%%%%%%%%%%%%%%%%%%%%%%%%%%%%%%%%%%%%%%%%%%%%
%%%%%
%%%%% Figure 6. Difference in v_imp
%%%%%
%\begin{figure*}
%  \centering
%  \includegraphics[width=\myfigwidthW\textwidth]{sfd_vaY.eps}
%  \caption{Dependence of impact velocity ($v_{\rm imp}$, km/s) on the angular
%   distance from apex ($\gamma$, degree) in our numerical result.
%   (a) For the population A. The solid line denotes the result of
%       linear least square fit, $v_{\rm imp} = 0.00300 \gamma + 19.0$.
%   (b) For the population B. The solid line denotes the result of
%       linear least square fit, $v_{\rm imp} = 0.00776 \gamma + 24.0$.
%  }
%  \label{fig:vr_diff}
%\end{figure*}

\begin{figure*}
  \centering
  \includegraphics[width=\myfigwidthW\textwidth]{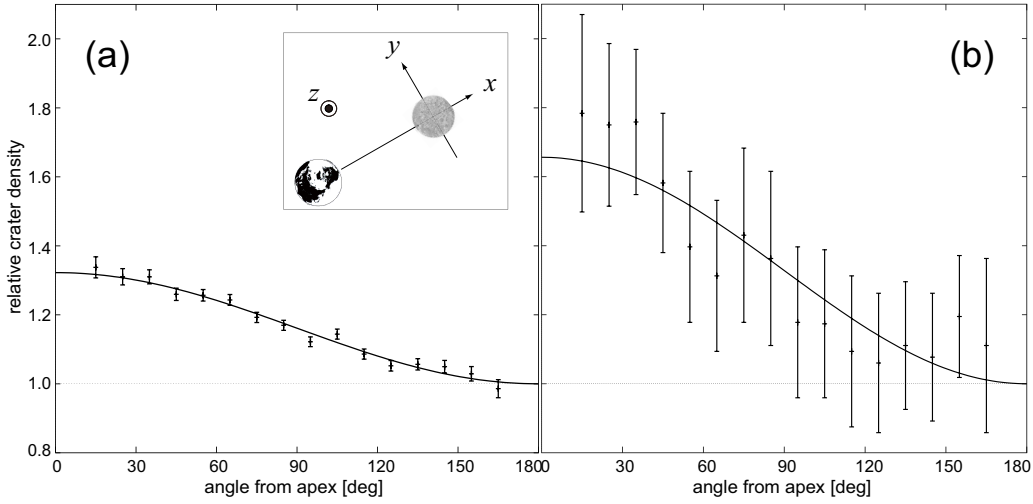}
  \caption{%
   Modeled and observed impact crater distribution on the Moon.
   We normalized the crater density to unity at antapex
   $(\gamma = 180^\circ)$ using the best fit sinusoid (solid line curve).
  (a) Numerical result including the correction due to the difference of
      average impact velocity
      as a function of the angular distance $(\gamma)$ from apex $(\gamma=0)$.
  (b) The observed rayed crater distribution of $D>5$ km \citep{morota2003}.
      The inset in (a) illustrates the coordinate system in this frame:
      The Earth always lies along $-x$ direction,
      the Moon velocity is toward $+y$ direction, and
      $+z$ is the north of the Earth--Moon system.
      Apex point is defined as $(x,y,z)=(0,R_{\rm M},0)$
      where $R_{\rm M}$ is the lunar radius. 
  }
  \label{fig:asymmetry}
\end{figure*}

Including this correction to our second stage simulation,
we computed the simulated spatial density of NEA
impacts on the Moon. Normalizing to unity at antapex, our
simulation result for the crater density as a function of apex
angle are shown in Fig.~\ref{fig:asymmetry}, panel (a).
In Fig.~\ref{fig:asymmetry}, we used a simple sinusoid with the
function form of $A + B \cos \gamma$ for a fitting curve where
$A$ and $B$ are fitting parameters, normalizing $A + B \cos 180^\circ = 1$.

For comparison, panel (b) 
shows the distribution found from the analysis of observed lunar
rayed craters \citep{morota2003}.
Note that the number of the lunar rayed craters in the observational
data analyzed by \citep{morota2003} is only 222,
while we have about 74,000 impacts in our simulation.
This difference is reflected in the difference of
the errorbar magnitudes in Fig.~\ref{fig:asymmetry}, which are based
on Poisson statistics.

Examining Fig.~\ref{fig:asymmetry},
what we notice first is that the apex/antapex asymmetry is less prominent
in the numerical results (panel (a)) compared with
the observed lunar rayed crater record (panel (b)).
The maximum crater density at apex is about 1.65
(normalized to unity at antapex, and estimated from the best-fit
sinusoid) in the observed crater record, whereas in our simulations,
it is
% 1.40  for the population A, and it is even smaller, about
% 1.32, for the population B particles and
$1.32 \pm 0.01$.

%%%%%
%%%%% Figure 6 (= originally Figure 7).
%%%%% Asymmetric distribution of craters/impacts along y-direction
%%%%%

\begin{figure*}
  \centering
  \includegraphics[width=\myfigwidthN\textwidth]{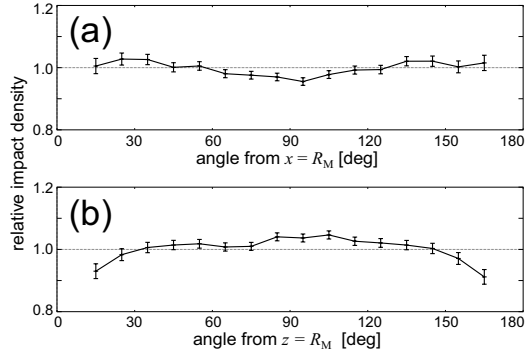}
  \caption{
   Relative impact density (impact flux) on the Moon in our numerical results
   (a) along $x$-axis with respect to the point $(R_{\rm M}, 0, 0)$, and
   (b) along $z$-axis with respect to the point $(0, 0, R_{\rm M})$.
   The average impact density over the range of angles
   $0$ to $180^\circ$ (from $x=R_{\rm M}$ for the $x$ direction and
                       from $z=R_{\rm M}$ for the $z$ direction)
   is normalized to unity.
   Note that the normalization here is different
   from that in Fig.~\protect{\ref{fig:asymmetry}}.
  }
  \label{fig:asymmetryXZ}
\end{figure*}

We also examined our numerical model result for trends in the NEA
impact density (impact flux) with respect to the angles along
the $x$ and $z$ axes (Fig.~\ref{fig:asymmetryXZ}).
Here we again adopt Poisson statistics to assign uncertainties in
the data plotted in Fig.~\ref{fig:asymmetryXZ}.
We find a tiny dip around the angle $90^\circ$ 
along the $x$ direction (Fig.~\ref{fig:asymmetryXZ}(a)),
although the numerical noise is large.
We can interpret this dip as related to the dip
that we see in the encounter statistics of particles at Earth's
activity sphere along the $x$ axis (cf.~Fig.~\ref{fig:rI_stat}(b))
and as owing to the same dynamical reason.

Along the $z$ axis, we notice a lower impact density at the polar regions
(Fig.~\ref{fig:asymmetryXZ}(b)).
We interpret this pattern as related to the encounter density
at the Earth's activity sphere along the $z$ direction,
Fig.~\ref{fig:rI_stat}(b).
That the number of particle encounters at Earth's activity sphere
becomes the smallest around the angle $= 90^\circ$
is reflected in the trend found in Fig.~\ref{fig:asymmetryXZ}(b).
The difference in cratering rate between at the polar and the equatorial
regions is $\sim10${\%};
this is consistent with
the analytical estimate by \citet{lefeuvre2008},
as well as the numerical result presented in \citet{gallant2009}.

%%%%%
%%%%% Asymmetric distribution of craters/impacts along the x- direction
%%%%%

%%%%%%%%%%%%%%%%%%%%%%%%%%%%%%%%%%%%%%%%%%%%%%%%%%%%%%%%%%%%%%%%%%%%%%%%%%%%
\section{Discussion\label{sec:discussion}}

Does the dynamical model of NEA impacts account for the observed asymmetry
of lunar rayed craters?  The observed crater record has relatively
large errorbars compared to our dynamical model results
(Fig.~\ref{fig:asymmetry}), and the apex/antapex
contrast of the observed and modeled crater densities may be considered
at least qualitatively consistent with the observations.  
Such a conclusion would imply that the young lunar craters are owed to
impacts of the NEAs whose orbital distribution we already know.

However, the intriguing systematic difference between the
degree of asymmetry between our dynamical model
and the observed crater record, though not enormous, is worthy of comment.
If we use the best-fit sinusoids in Fig.~\ref{fig:asymmetry}, we can say
that the dynamical model accounts for only about
49{\%} $(=(1.32-1)/(1.65-1))$
%%% 79{\%} $(=1.32/1.65)$
of the observed lunar rayed crater asymmetry.
We would like to pursue the reasons for this potential discrepancy,
hoping to understand and constrain better the dynamical origin of the projectiles that have
created the lunar rayed craters over the past $\sim$one billion years.
Because the uncertainties in our numerical model are much smaller than those
in the observational data (compare the errorbars in Fig.~\ref{fig:asymmetry}a
and Fig.~\ref{fig:asymmetry}b), we must be very cautious in the
discussion of this point.  
We emphasize that the possible explanations discussed 
below must await assessment with improved observational data.

One possible explanation is related to the impact velocity distributions.
The leading/trailing asymmetry becomes more prominent when the average
relative velocity between the Moon and the projectiles is low.
The NEA-like particles are, by their dynamical definition,
the ``slowest'' (relative to Earth) 
among all the known small body populations in the solar system.
That even these slow particles may not fully account for
the observed asymmetric distribution 
in the lunar crater record suggests that there may exist a
presently-unobserved 
population of small objects near the Earth's orbit that have even lower
average relative velocity than the currently known near-Earth asteroids do.
Conventional debiasing techniques, such as in \citet{bottke2000,bottke2002},
would not enhance the NEA-like particles with low relative velocity.
Rather, such debiasing generally compensates for the existence of more distant
objects with higher relative velocity (i.e., larger $e,I$).

This argument predicts the existence of a hitherto unseen population of
slow objects whose heliocentric orbits are close to the Earth--Moon
system. We make a rough estimate of the unseen population as follows.
The best-fit sinusoid to the observational lunar crater asymmetry
is consistent with an impactor population with average lunar impact velocity of
10--12 km/s \citep{morota2003}.
Consider an impactor population having a similar shape
of the impact velocity distribution function as the simulated NEAs 
(Fig.~\ref{fig:rM_stat1}a) but with
$\langle v_{\rm imp} \rangle = \mbox{10--12}$ km/s rather than the
$\sim$22 km/s that we found in our dynamical model based on the known NEAs.
In such a population, the fraction of slow objects would be roughly 50{\%}
more than the fraction of slow objects in the currently known population of
NEAs; here we define ``slow'' NEAs as those having potential lunar
impact velocity $< 12$ km/s; such objects would be nearly coorbiting
with Earth. In other words, our rough estimate of the actual slow NEA
population is $\sim50{\%}$ more than the known slow NEAs.

Such a population could have remained undetected in observational
surveys to date either 
because the surveys have low sensitivity to their orbital parameters
or because these objects are fainter (smaller and/or darker).
The rayed crater record in Fig.~\ref{fig:asymmetry}(b) contains craters
with diameter $D > 5$ km. On the lunar surface, a crater with $D \approx 5$ km
can be created by an asteroidal projectile with
$D_{\rm projectile} =$ 0.2--0.3 km even when
the impact velocity is as small as 10 km/s and when the projectile density
is that of porous rock ($\sim$ 1.5 g/cm${}^3$).
These small and slow objects, if they exist in the greater numbers that our
study indicates,
could account for the discrepancy between our numerical result
and the observed asymmetric crater distribution.
More complete observational surveys of
the near-Earth asteroids can test our prediction.
Future progress in the reconstruction of the true
orbital distribution of NEAs by debiasing techniques would also be useful.

However, a challenge with the above explanation is that it is not easy to 
keep NEAs' relative velocity too low.
There are many complicated resonances in the orbital zones of the
terrestrial planets
that can pump up the random velocities of small bodies \citep{michel97}.
Thus the slow population would need to be immune to these
excitation mechanisms or to be continuously resupplied.

A different explanation might be that larger NEAs with very
low relative velocity get fragmented due to Earth's tidal force when 
they approach the Earth--Moon system.
This process would increase the number of 
projectiles (and reduce their size), and may contribute in 
enhancing the asymmetric distribution of craters if the fragments keep the low
relative velocity of the parent body until they collide with the Moon.

A third possibility is that 
the lunar orbital velocity has been larger in the past.
A billion years ago the lunar semimajor axis may have been as small as
$\sim$ 90{\%} of the current value \citep{bills99},
and has gradually increased to the current value
due to the tidal interaction with the Earth.
When the lunar semimajor axis was 90{\%} of the current value,
the lunar orbital velocity with respect to the Earth was 17{\%}
larger than the current value.
This enhancement of the lunar orbital velocity could enhance the
asymmetric distribution of impacts.
But the magnitude of this effect would be limited.
Even if the lunar orbital velocity has been 17{\%} larger throughout
the past one billion years, it would be only as small as 1.2 km/s, still too
small compared to the average impact velocity.
This larger value would still be insufficient to explain the difference of
the degree of asymmetric distribution of the actual lunar craters record
(apex/antapex ratio $\sim 1.65$) from that of our numerical result
($\sim$1.32) when we consider the semi-analytic estimate of the cratering
rate as a function of the lunar orbital velocity \citep{zahnle2001,morota2003}.
Similarly, \citet{gallant2009} performed a series of numerical simulations
in order to check the effect of the smaller Earth--Moon distance, and found 
only a tiny change in hemispherical crater ratio for Earth--Moon
distance as low as $\sim$90{\%} of the current value
($\sim$54 Earth radii).
We must note, however, that the history of the lunar orbit is thus far
predominantly based on theoretical models and is not especially well 
constrained by observations;  there may exist an exciting possibility
to place an observational constraint on the
lunar orbital evolution by detailed modeling of the asymmetric
lunar crater record.

A fourth possibility is that the observational lunar crater data of
\citet{morota2003} is incomplete. This crater data consists only of 222
craters of diameter $D>5$ km
covering about a third of the entire lunar surface.  There is certainly
room for improvement of this dataset. A recent brief report of a more
extensive search for lunar rayed craters as small as 0.5 km diameter is not
conclusive \citep{werner2010}. A potentially important source of
uncertainty and confusion in interpreting the spatial patterns in the
lunar rayed craters is the uncertain
ages of these craters.  For example, some of the larger craters are argued
to be older than what they had been thought from a study of optical
maturity of their ejecta \citep{grier2001}.  Similar issues are pointed
out by \citet{werner2010}.  Thus, it is possible that the discrepancy
could be removed with a future complete and correct dataset  of young
lunar craters.

Finally, we should comment on the study of
\citet{gallant2009} which already reported results that are overall
rather similar to ours.  
The first and the largest difference of our numerical model from
that of \citet{gallant2009} is that the latter 
simulated the lunar impacts of NEAs from a synthetic NEA
sample having fixed orbital elements with distribution following 
\citet{bottke2002}'s debiased NEA population
(referred to as ``source orbits'', restricted to 16,307 in
the Earth-crossing region), whereas
we directly integrated the orbital evolution of Bottke's debiased NEA
population particles (in our first stage simulation) with a
steady-state dynamical model. Then, using symmetric characteristics
of the orbits, \citet{gallant2009} effectively multiplied the number
of source orbits four times. Next, for each of the source orbits,
a disk of $10^5$ particles with an identical initial velocity toward
the Earth--Moon system was created, and integrated until the disk
particles reach the Earth.
Schematic figure of Fig.~4 in \citet{gallant2009} explains well their
numerical model of the disk; the total number of particles in
their numerical model is more than $1.2 \times 10^{11}$.
In comparison, our model contains a smaller number of particles 
($0.21 \times 10^{11}$ clones),
but we have included planetary perturbations and orbital evolution
in direct numerical integrations of the NEA-like particles.
%%%
%%% NOTE: It is not yet too late to increase the number of clones in our
%%% numerical model so that it becomes larger than \citet{gallant2009} $\cdots$
%%% particularly if I can use some part of your cluster as well.
%%% It will take another month, but would it be worth?

\citet{gallant2009} concluded that their numerical result of the apex/antapex
asymmetry ($1.28 \pm 0.01$ when considering the ratio of craters
within $30^\circ$ of the apex to those
within $30^\circ$ of the antapex) is completely consistent with the
value of about $1.6 \pm 0.1$
found in the available crater data of \citet{morota2003};
they attributed the difference to the large uncertainties in the crater
data.    As we noted above, the difference between the dynamical model and the observations
is not huge, but, somewhat differently than \citet{gallant2009}, we conclude that the the results
of our numerical simulation are only marginally consistent with the observed asymmetry, and we
have therefore discussed at some length several explanations for the possible discrepancy. 

\citet{gallant2009} also pointed out that the
impact velocity of the cratering projectiles is $\sim20$ km/s,
somewhat higher than values commonly adopted in previous
studies, and that this has ramifications for proposed matches between 
the lunar crater size-frequency distributions and asteroidal impactors 
\citep[e.g.][]{strom2005}.  Our calculations find the average
impact velocity to be 22.4 km/s, which is
even slightly higher than that of \citet{gallant2009}.
Not only the average impact velocity, but also the shape of the
impact velocity distribution in our model is noticeably different than 
in \citet{gallant2009}: it is more symmetric about the mean value
in our case.
\textcolor{\mynotecolor}{
We attribute these differences partly to the 
statistical variations of the initial conditions (both studies used 
Bottke et al.~(2002)'s debiased $a,e,I$ distribution of NEO orbits, but
the particular realizations of the set of initial conditions were done 
independently), and partly to the evolution of the particles prior to
impact on the Moon (\citet{gallant2009}'s numerical model does not 
account for this orbital evolution, but our dynamical model includes 
this effect).
}
Note that even though
the average impact velocity as well as the shape of the impact velocity
distribution in our model are different from those in \citet{gallant2009},
the resulting asymmetry in the numerical lunar cratering is quite
similar to each other.

Regarding implications for origins of crater populations, certainly the higher
average impact velocity of NEAs on the Moon calls for an update of such
studies, but with the caveat that this higher value is derived from the
currently recognized ``debiased" NEA population (which may potentially be 
missing a significant fraction of slow NEAs near the Earth-Moon system).   
We also note that the Late Heavy Bombardment projectiles 
that \citet{strom2005} proposed
were not NEAs with the steady-state flux but main belt asteroids directly 
transported from main belt resonance zones to the inner solar system; 
such impactors would have an average impact velocity on the Moon 
of about 18 km/s \citep{ito2006},
quite similar to the value of 17 km/s adopted in the \citet{strom2005} study.

Currently several lunar missions are underway
by several countries \citep[e.g. ][]{normille2007}.
They will yield improved datasets to better determine the nature
of the asymmetric distribution of young craters on the Moon.
On the theoretical side, it would be important to improve the
dynamical models by including more complete physics (such as
non-gravitational forces that may be significant in the orbital
evolution of small NEAs), and to improve the model estimates of 
observational biases in the known NEA population particularly
for those with orbital parameters similar to Earth and that  are difficult
to observe due to their low solar elongation angles.

%%%%%%%%%%%%%%%%%%%%%%%%%%%%%%%%%%%%%%%%%%%%%%%%%%%%%%%%%%%%%%%%%%%%%%%%%%%%
\textcolor{\mynotecolor}{%
\section{Summary and Conclusions\label{sec:conclusions}}
}

\textcolor{\mynotecolor}{%
We simulated numerically the spatial distribution of impacts of near-Earth
objects, using a numerical model with a steady-state population of
impactors based on current estimates of the debiased near-Earth objects'
orbital distribution (as provided by the model of \citet{bottke2002}).
We compared the results of the simulation with the observed asymmetry of
the population of rayed craters on the leading/trailing hemispheres of the
Moon (as measured by \citet{morota2003}).
Our results and conclusions are summarized as follows.
}

\textcolor{\mynotecolor}{%
\begin{enumerate}
\item{} Our numerical simulation yields a leading/trailing hemispherical
  ratio of $1.32\pm0.01$ for lunar impacts by near-Earth objects.  This
  result is similar to that of \citet{gallant2009} who obtained the value
  $1.28\pm0.01$ from a different numerical model.
\item{} This result of our numerical simulation is only marginally
  compatible with the observed ratio of $1.65 \pm 0.16$ found by
  \citet{morota2003}.  A possible explanation for the discrepancy is that
  there exists a hitherto undetected population of small objects in
  heliocentric orbits nearly coorbiting with Earth, whose average impact
  velocities on the Moon are much lower than the average impact velocity of
  the known near-Earth object population.  Other explanations are possible,
  including the possibility that a more comprehensive study of young lunar
  craters could reveal a smaller leading/trailing asymmetry and thereby
  remove the discrepancy with the dynamical modeling.
\item{} The average impact velocity of near-Earth objects on the Moon is
  found to be 22.4 km/s; Fig.~\ref{fig:rM_stat1} plots the impact
  velocity distribution.
\item{} Overall, the impact angles are isotropically distributed, but
  there is a noticeable deficit of oblique impacts on the trailing
  hemisphere of the Moon.
\item{} The ratio of the number of collisions with the Earth and those 
  with the Moon is found to be $20.4 \pm 0.1$.
\end{enumerate}
}

%%%%%%%%%%%%%%%%%%%%%%%%%%%%%%%%%%%%%%%%%%%%%%%%%%%%%%%%%%%%%%%%%%%%%%%%%%%%%%
\begin{acknowledgements}
We wish to thank Tomokatsu Morota, Fumi Yoshida,
and Robert McMillan for giving us useful
information and discussions.
% Detailed and constructive review by Yolande McLean has considerably
% improved the presentation of this paper.
% We wish to thank William Bottke for many comments
% and suggestions that improved this paper.
We wish to thank the reviewer Bill Bottke for many helpful comments
and suggestions that led to an improved numerical model.
This study is supported by the Grant-in-Aid of the Ministry of Education
of Japan (18540426/2006--2008, 21540442/2009--2011) and
the JSPS program for Asia--Africa academic platform (2009--2011).
RM acknowledges research funding from the USA--NSF grant (AST--0806828).
\end{acknowledgements}

%%%
%%% References (from Bib files)
%%%
%\bibliographystyle{aa}
%\bibliography{mybib} % {myown,tanikawa}

\begin{thebibliography}{26}
\expandafter\ifx\csname natexlab\endcsname\relax\def\natexlab#1{#1}\fi

\bibitem[{Bills {et~al.}(1999)Bills, James, \& Mengel}]{bills99}
Bills, {\mbox{B.G}}., James, {\mbox{T.S}}., \& Mengel, {\mbox{J.G}}. 1999, J.
  Geophys. Res., 104, 1059

\bibitem[{Bottke {et~al.}(2000)Bottke, Jedicke, Morbidelli, Vokrouhlick\'y,
  Bro\v{z}, Nesvorn\'y, Petit, \& Gladman}]{bottke2000}
Bottke, {\mbox{W.F}}., Jedicke, R., Morbidelli, A., {et~al.} 2000, Science,
  288, 2190

\bibitem[{Bottke {et~al.}(2002)Bottke, Morbidelli, Jedicke, Petit, Levison,
  Michel, \& Metcalfe}]{bottke2002}
Bottke, {\mbox{W.F}}., Morbidelli, A., Jedicke, R., {et~al.} 2002, Icarus, 156,
  399

\bibitem[{Danby(1992)}]{danby92}
Danby, {\mbox{J.M.A}}. 1992, Fundamentals of Celestial Mechanics (Richmond,
  Virginia: Willmann--Bell Inc.), second edition

\bibitem[{Gallant {et~al.}(2009)Gallant, Gladman, \& \'Cuk}]{gallant2009}
Gallant, J., Gladman, B., \& \'Cuk, M. 2009, Icarus, 202, 371

\bibitem[{Grier {et~al.}(2001)Grier, {\mbox{McEwen}}, Lucey, Milazzo, \&
  Strom}]{grier2001}
Grier, {\mbox{J.A}}., {\mbox{McEwen}}, {\mbox{A.S}}., Lucey, {\mbox{P.G}}.,
  Milazzo, M., \& Strom, {\mbox{R.G}}. 2001, J. Geophys. Res., 106, 32847

\bibitem[{Housen {et~al.}(1991)Housen, Schmidt, \& Holsapple}]{housen91}
Housen, {\mbox{K.R}}., Schmidt, {\mbox{R.M}}., \& Holsapple, {\mbox{K.A}}.
  1991, Icarus, 94, 180

\bibitem[{Ishizaki \& Furumoto(1997)}]{ishizaki97e}
Ishizaki, Y. \& Furumoto, M. 1997, Planet. People, 6, 12, text in Japanese

\bibitem[{Ito \& Malhotra(2006)}]{ito2006}
Ito, T. \& Malhotra, R. 2006, Adv. Space Res., 38, 817

\bibitem[{Ivanov {et~al.}(2002)Ivanov, Neukum, Bottke, \&
  Hartmann}]{ivanov2002}
Ivanov, {\mbox{B.A}}., Neukum, G., Bottke, {\mbox{W.F}}., \& Hartmann,
  {\mbox{W.K}}. 2002, in Asteroids III, ed. {\mbox{W.F}}.~Bottke, A.~Cellino,
  P.~Paolicchi, \& {\mbox{R.P}}.~Binzel (Tucson, Arizona: The University of
  Arizona Press), 603--612

\bibitem[{{\mbox{Le Feuvre}} \& Wieczorek(2008)}]{lefeuvre2008}
{\mbox{Le Feuvre}}, M. \& Wieczorek, {\mbox{M.A}}. 2008, Icarus, 197, 291

\bibitem[{Levison \& Duncan(1994)}]{levison94}
Levison, {\mbox{H.F}}. \& Duncan, {\mbox{M.J}}. 1994, Icarus, 108, 18

\bibitem[{McEwen {et~al.}(1997)McEwen, Moore, \& Shoemaker}]{mcewen97}
McEwen, {\mbox{A.S}}., Moore, {\mbox{J.M}}., \& Shoemaker, {\mbox{E.M}}. 1997,
  J. Geophys. Res., 102, 9231

\bibitem[{Melosh(1989)}]{melosh89}
Melosh, {\mbox{H.J}}. 1989, Oxford Monographs on Geology and Geophysics,
  Vol.~11, Impact Cratering: A Geologic Process (New York: Oxford University
  Press)

\bibitem[{Michel \& Froeschl\'e(1997)}]{michel97}
Michel, P. \& Froeschl\'e, C. 1997, Icarus, 128, 230

\bibitem[{Morota \& Furumoto(2003)}]{morota2003}
Morota, T. \& Furumoto, M. 2003, Earth Planet. Sci. Lett., 206, 315

\bibitem[{Normille \& Bagla(2007)}]{normille2007}
Normille, D. \& Bagla, P. 2007, Science, 317, 1163

\bibitem[{Schenk \& Sobieszczyk(1999)}]{schenk99}
Schenk, P. \& Sobieszczyk, S. 1999, Bull. Am. Astron. Soc., 31, 1182

\bibitem[{Schmidt \& Housen(1987)}]{schmidt87}
Schmidt, {\mbox{R.M}}. \& Housen, {\mbox{K.R}}. 1987, Int. J. Impact Eng., 5,
  543

\bibitem[{Shoemaker(1962)}]{shoemaker62}
Shoemaker, {\mbox{E.M}}. 1962, in Physics and Astronomy of the Moon (New York:
  Academic Press), 283--359

\bibitem[{Shoemaker {et~al.}(1982)Shoemaker, Lucchitta, Wilhelms, Plescia, \&
  Squyres}]{shoemaker82}
Shoemaker, {\mbox{E.M}}., Lucchitta, {\mbox{B.K}}., Wilhelms, {\mbox{D.E}}.,
  Plescia, {\mbox{J.B}}., \& Squyres, {\mbox{S.W}}. 1982, in Satellites of
  Jupiter, ed. D.~Morrison (Tucson: The University of Arizona Press), 435--520

\bibitem[{Strom {et~al.}(2005)Strom, Malhotra, Ito, Yoshida, \&
  Kring}]{strom2005}
Strom, {\mbox{R.G}}., Malhotra, R., Ito, T., Yoshida, F., \& Kring,
  {\mbox{D.A}}. 2005, Science, 309, 1847

\bibitem[{Werner \& Medvedev(2010)}]{werner2010}
Werner, {\mbox{S.C.}}. \& Medvedev, S. 2010, Lunar Planet. Sci. Conf., 41, 1058

\bibitem[{Zahnle {et~al.}(1998)Zahnle, Dones, \& Levison}]{zahnle98}
Zahnle, {\mbox{K.J}}., Dones, L., \& Levison, {\mbox{H.F}}. 1998, Icarus, 136,
  202

\bibitem[{Zahnle {et~al.}(2001)Zahnle, Schenk, Sobieszczyk, Dones, \&
  Levison}]{zahnle2001}
Zahnle, {\mbox{K.J}}., Schenk, P., Sobieszczyk, S., Dones, L., \& Levison,
  {\mbox{H.F}}. 2001, Icarus, 153, 111

\bibitem[{Zahnle \& Sleep(1997)}]{zahnle97}
Zahnle, {\mbox{K,J}}. \& Sleep, {\mbox{N.H}}. 1997, in Comets and the Origin
  and Evolution of Life, ed. {\mbox{P.J}}.~Thomas, {\mbox{C.F}}.~Chyba, \&
  {\mbox{C.P}}.~McKay (New York: Springer--Verlag), 175--208

\end{thebibliography}

%\end{document}

%%%
%%% References (for submission)
%%%

\end{document}